\documentclass[12pt,showpacs,amsmath,amssymb]{revtex4}
\usepackage{bm}
\begin{document}
\title{Decay of polarized $\Delta$}
\author{G Ramachandran$^1$, Venkataraya$^{1,2}$, M S Vidya$^1$, \\ J Balasubramanyam$^{1,3}$ and G Padmanabha$^{1,4,5}$}
\address{$^1$ GVK Academy, Jayanagar, Bangalore - 560082, India}
\address{$^2$ Vijaya College, Bangalore - 560 011, India}
\address{$^3$ K. S. Institute of Technology, Bangalore - 560 062 , India}
\address{$^4$ Department of Physics, Bangalore University, Bangalore - 560 056, India}
\address{$^5$ SBM Jain College, V V Puram, Bangalore - 560004, India}

\begin{abstract}
\vspace{0.5in}
The resonance $\Delta(1232)$ with spin-parity ${3 \over 2}^+$, which contributes dominantly to the reactions
like $\gamma N \to \pi N$ and $NN \to NN\pi$ at intermediate energies, may be expected to be produced 
in characteristically different polarized spin states. As such an analysis of the decay of polarized delta is presented,
which may be utilized to probe empirically  the production mechanism. It is shown that 
measurements of the angular distributions of the pion and the polarization of the nucleon arising out of  
${\vec{\Delta}}$ decay can determine empirically the Fano statistical tensors 
characterizing the ${\vec{\Delta}}$. 

\end{abstract}

\pacs{25.20.Ljs,24.70.+s,14.20.Gk ,13.60.Le,13.30.-a}

\maketitle
\vspace{0.5in}
Meson production in $NN$ collisions~\cite{mach} as well as photo- and electro-production of 
mesons \cite{krush} have attracted renewed interest with the technological 
advances ~\cite{mach,krush,pollock} made at  IUCF Indiana in 
USA, the development of COSY at Julich in Germany and 
 the advent of new generation of electron
accelerators at J-Lab, MIT, BNL in USA, ELSA at Bonn, MAMI at Mainz in Germany, ESRF at Grenoble
in France and Spring$8$ at Osaka in Japan. The excitement began in the early $1990$'s, when
the total cross section measurements ~\cite{meyer} for $pp\rightarrow pp\pi^0$ were found to 
exceed the then available theoretical predictions ~\cite{koltan}, by more than a factor of $5$.
Moreover, a large momentum transfer is involved when an additional particle is produced in the 
final state. This implies that the features of the $NN$ interaction is probed at very short distances,
estimated ~\cite{nakayama} to be of the order of $0.53$ fm, $0.21$ fm and $0.18$ fm for the production of
$\pi$, $\omega$ and $\phi$ mesons respectively. While $\omega$ and $\phi$ production involve 
only excited nucleon states, pion production involves $\Delta$ which is a 
well isolated resonance. Moreover, experimental studies on pion production 
in $pp$ collisions have reached a high degree of sophistication ~\cite{prez, meyer1}, 
where both the protons are polarized initially and the three body final state 
is completely identified kinematically. The
Julich meson exchange model ~\cite{hanhart}, which yielded theoretical predictions closer to the
experimental data than most other models, has been more successful in the case of charged pion
production ~\cite{prez} than with the production of neutral pions ~\cite{meyer1}. 
A more recent analysis ~\cite{deepak}
of the ${\vec p}{\vec p} \rightarrow pp \pi^0$ measurements ~\cite{meyer1}, following a model 
independent irreducible tensor approach ~\cite{gr}, showed that the Julich model deviates quite 
significantly for the $^3P_1 \rightarrow{^3P}_0 p$ and to a lesser extent for the $^3F_3 \rightarrow{^3}P_2 p$
transitions. These calculations ~\cite{deepak} have been carried out with and without taking 
the $\Delta$ contributions into account and this exercise emphasized the important role of $\Delta$ in the
model calculations. It has been shown ~\cite{silbar} that $NN\rightarrow N\Delta$ reaction matrix contains as many 
as $16$ amplitudes out of which $10$ are second rank tensors. Ray ~\cite{ray} has drawn attention 
to their importance based on a partial wave expansion model
where he found that ``{\it the total and differential cross-section reduced by about one half, the structure
in the analyzing powers increased dramatically, the predictions of $D_{NN}$ became much too
negative, while that for $D_{LL}$ became much too positive and the spin correlation predictions were
much too small when all ten of the rank 2 tensor amplitudes were set to zero, while the remaining
six amplitudes were unchanged} ". The dominance of $\Delta$
in photo and electro-pion production has been known for a long time. 
Very recently, a theoretical formalism ~\cite{photopapers} for photo and electroproduction 
of mesons with arbitrary spin-parity $s^\pi$
has been outlined, which specializes to pion production for $s^\pi = 0^-$. 
The  excitation of a nucleon into $\Delta$ through photo-absorption, 
involves electric quadrupole $(E2)$ and magnetic dipole $(M1)$ form factors, while electro-production involves a Coulomb form factor 
$(C2)$ in addition. The electro-excitation of $\Delta$ provides a test~\cite{carlson} 
for perturbative QCD, which appears to fail when confronted with experiments~\cite{beck}.
The study of these form factors as well as the quadrupole deformation of $\Delta$ has excited
considerable attention~\cite{arhens}. The importance of $\Delta$ has also been highlighted in more recent studies \cite{exp}. 

One can naturally expect $\Delta$, whether it is produced through electromagnetic 
excitation or hadronic excitation, to be polarized. Characterizing the $\vec{\Delta}$ produced 
by the Fano~\cite{fano} statistical tensors $t^k_q$
of rank $k = 1,2,3$ with $q = -k, -k+1, ....., k$, we discuss here the extent of information that can be 
obtained on the $t^k_q$ through measurements of the angular distribution of the pion and that of the nucleon polarization in the final state.

 For this purpose, we express the reaction matrix $M$ for the decay of $\Delta$ in the form 
 \begin{equation}
 M = f (S^1({\textstyle \frac{1}{2},\frac{3}{2}})\cdot Y_{1} ({\bf \hat q})),
 \end{equation}
 where ${\bf q}$ denotes the meson momentum in the $\Delta$ rest frame and ${\bf \hat q}= \bf q / |{\bf q}|$. 
The transition spin operators $S^\lambda _m (s_f,s_i)$ of rank $\lambda$, connecting the initial and final
channel spins $s_i$ and $s_f$ respectively, are defined following~\cite{gr2}, the multiplicative factor $f$
defines the strength of the transition and $Y_{l,m}({\bf \hat q})$ denote spherical harmonics. The state
of polarization of $\Delta$ is, in general, represented by the spin density matrix
\begin{equation}
\rho^{\Delta} = \frac{Tr(\rho^\Delta)}{4} \sum_{k = 0}^3 (\tau^k \cdot t^k),
\end{equation} 
in terms of Fano statistical tensors 
\begin{equation}
t^k_q = \frac{Tr(\tau^k_q~\rho^\Delta)}{Tr{\rho^\Delta}},
\end{equation}
where $\tau^k_q$ denote irreducible tensor operators of rank $k$ in the $\Delta$ spin space
of dimension $2j+1 = 4$. The hermitian conjugates $(\tau^k_q)^\dagger$ are related to $(\tau^k_q)$  through
$(\tau^k_q)^\dagger = (-1)^q \tau^k_{-q}$ and are so normalized as to satisfy 
\begin{equation}
Tr[\tau^k_q \,(\tau^{k'}_{q'})^\dagger] = (2j+1) \,\delta_{kk'} \,\,\delta_{qq'},
\label{tau}
\end{equation}
with $\tau^0_0 = 1$, so that $t^0_0 = 1$. The above normalization is different 
from that used by Fano originally, but consistent with
that chosen~\cite{madison} for spin $j = 1$. 
It may be noted that $\tau^1_q$ for spin-$3/2$ are related to the
spherical components  
$
 J^1_0 =  J_z \nonumber \; ; \;
 J^1_{\pm 1} = \frac {\mp(J_x \pm i J_y)}{{\sqrt 2}} 
$
of the spin operator ${\boldsymbol J}$ of $\Delta$ 
 through 
 \begin{equation}
 \tau^1_q = {\sqrt \frac{4}{ 5}} \,J^1_q,
 \end{equation}
 whereas of $\tau^1_q = {\sqrt \frac{3}{ 2}}J^1_q$ in the case of spin $j=1$.

  Our purpose is to determine empirically  the $t^k_q$ for $k = 1,2,3$ of $\vec \Delta$ through an 
  experimental study of its decay products  
 
  The angular distribution of the pion in the case of  $\vec \Delta$ decay is given by 
\begin{equation}
I({\bf \hat q}) = Tr(M\,\rho^{\Delta}\,M^\dagger),
\label{iq}
\end{equation}
where $M^\dagger$ denotes the hermitian conjugate of $M$. Noting 
that $\tau^k_q \equiv S^k_q({\textstyle \frac{3}{2},\frac{3}{2}})$ and making
use of the known properties~\cite{gr2} of the spin operators, we obtain
\begin{equation}
I({\bf \hat q}) = \sum_{k=0,2}\,\, \sum_{q=-k}^{k}(-1)^q\, I^k_{-q} \, Y_{kq}({\bf \hat q}),  
\end{equation}
in terms of 
\begin{equation}
I^k_q = \frac{9\,|f|^2}{4\,\sqrt \pi}\,{\textstyle W(\frac{3}{2}k\frac{1}{2}1;\frac{3}{2}1)}\,C(1k1;000)\,t^k_q \, ,
\end{equation}
which may be measured experimentally, since $Y_{kq}({\bf \hat q})$ are orthonormal to each other. 
Note that $t^0_0 = 1$, yields $|f|^2$ empirically from an experimental measurement of $I^0_0$. 
This information may then be used to determine $t^2_q$ from experimentally 
measured $I^2_q$. We next observe that the spin density matrix $\rho^N$ of the nucleon
in the final state is given by 
\begin{equation}
\rho^N = M\,\rho^\Delta \,M^\dagger,
\end{equation}
which may be expressed in the standard form
\begin{equation}
\rho^N = \frac{Tr({\rho^N})}{2} \,[1 + {\boldsymbol \sigma}\cdot{\bf P}],
\end{equation}
in terms of the Pauli spin matrices ${\boldsymbol \sigma}$ of the nucleon and nucleon polarization 
\begin{equation}
{\bf P} = \frac{Tr({\boldsymbol \sigma}\,\rho^N)} {Tr (\rho^N)},
\end{equation}
where the denominator is already known from eq.(\ref{iq}) as $I({\bf \hat q})$. 
Expressing $({\boldsymbol \sigma} . {\bf P})$ in terms
of  spherical components through $({\boldsymbol \sigma}\cdot {\bf P}) 
= \sum_\nu (-1)^\nu \,\sigma^1_{-\nu} P^1_\nu$ and noting
further that 
\begin{equation} 
\sigma^1_\nu \equiv S^1_\nu ({\textstyle \frac{1}{2},\frac{1}{2}}), 
\end{equation}
we obtain
\begin{eqnarray}
P^1_\nu({\bf \hat q}) &=& 9{\sqrt\frac{3}{2\pi}}\,|f|^2\,\, [I({\bf \hat q})]^{-1} 
\sum_{k=1}^3\,\sum_{q=-k}^k t^k_q \sum_{l=0,2} \, G(k,l)\,  \nonumber \\ 
& & \times \,C(kl1;qm\nu) \,Y_{lm}({\hat{\bf q}}),
\label{pol}
\end{eqnarray}
where the geometrical factors are 
\begin{eqnarray}
G(k,l)&=& \sum_{\Lambda} (-1)^{(k-\Lambda)} [k]\,[\Lambda]\,{\textstyle W(\frac{3}{2}k\frac{1}{2}1;\frac{3}{2}\Lambda)\,W(\frac{1}{2}1\frac{1}{2}\Lambda;\frac{3}{2}1)} \nonumber \\
& &\times W(k111;\Lambda l)\, C(11l;000)
\end{eqnarray}

We note that $I({\bf \hat q})$ and $|f|^2$ are already known, so that the spherically symmetric 
term with $l=0$ in eq.(\ref{pol}) yields $t^1_q$. The $l=2$ terms contain all the $t^k_q$ with 
$k = 1,2,3$. However, since it has already been shown that $t^2_q$ and $t^1_q$  are
determinable empirically from $I({\hat{\bf q}})$ and $l=0$ term of eq.(\ref{pol}), experimental study of the angular distribution with $l = 2$ may be 
used to determine $t^3_q$ empirically. 

One can expect the  empirical study of ${\Delta}$ polarization to yield useful information 
with regard to the formation of $\Delta$ in different scenarios. These detailed case by case studies will be taken up elsewhere. Apart from this, 
we may also draw attention to  some interesting aspects of $\Delta$ polarization. While $t^1_q$ determine 
the magnitude as well as the direction of the vector polarization, which is an axial vector, 
it is interesting to point out that two other independent axes are associated~\cite{vr} with the second 
rank polarization $t^2_q$, which define also the Principal Axes of Alignment. Hence the 
measurements  of the $t^2_q$ determine also the corresponding Principal Axes of Alignment Frame (PAAF) of $\vec{\Delta}$. 
Moreover, three additional independent axes are associated~\cite{vr1}  with the third rank tensor $t^3_q$, 
so that a maximal set of six independent axes characterize $\vec{\Delta}$, apart from 
the relative strengths of vector, second rank and third rank tensor polarizations. If the 
${\vec \Delta}$ produced is oriented ~\cite{ksm}, which is the simplest case of polarization of 
$\Delta$, all the six independent axes collapse into one. Such a scenario is quite unlikely especially in 
$NN \to NN\pi$. We, therefore, advocate the experimental study of the angular distribution of the pion 
and that of the nucleon polarization for  events, which correspond kinematically to invariant 
mass of the $\pi-N$ system being equal to the mass of the $\Delta$ resonance.
\vspace{0.3in}

\noindent
{\large{\bf Acknowledgements}}\\
Three of us (Venkataraya, JB and GP) thank the Principals and Managements of their respective
organizations for encouragement given to research work.

\end{document}